\documentclass[preprint,12pt]{elsarticle}
\usepackage{graphics}
\usepackage{array}

\usepackage{amssymb}
\usepackage{amsthm}

\journal{Physics Letters A}

\begin{document}

\begin{frontmatter}

\title{New Kochen-Specker Sets in Four Dimensions}

\author[label1]{Mladen Pavi\v ci\'c}

\author[label2]{Norman D.~Megill}

\author[label3]{Jean-Pierre Merlet}

\address[label1]{Chair of Physics, Faculty of Civil Engineering,
University of Zagreb, Croatia}

\address[label2]{Boston Information Group, 19 Locke Ln., Lexington,
MA 02420, U.~S.~A.}

\address[label3]{INRIA, projet COPRIN:
06902 Sophia Antipolis Cedex, France}

\begin{abstract}

We show that all possible 388 4-dim Kochen-Specker (KS) (vector)
sets (of yes-no questions) with 18 through 23 vectors and 844 
sets with 24 vectors all with component values from \{-1,0,1\} 
can be obtained by stripping vectors off a single system provided 
by  Peres 20 years ago. In addition to them, we have found a number 
of other KS sets with 22 through 24 vectors. We present the
algorithms we used and features we found, such as, for instance,
that Peres' 24-24 KS set has altogether six critical KS subsets.
\end{abstract}


\end{frontmatter}

\section{Introduction}
\label{sec:intro}

The Kochen-Specker (KS) theorem has recently been given renewed
attention due to new theoretical results which then prompted new 
experimental and computational techniques.

The new theoretical results concern the conditions under
which such experiments are feasible at all
\cite{meyer99,mermin99,cabell-02,barrett}, including a possible
way to formulate the KS theorem for single qubits
\cite{cabell-03,cabello-moreno-07}. Such results and
experiments enable applications in quantum computation
(restrictions imposed on complex configurations of
quantum gates, implementations of KS configurations of
quantum gates that rule out classical solutions, etc.).

The experiments were carried out for
spin$-\frac{1}{2}\otimes\frac{1}{2}$ particles (correlated
photons or spatial and spin neutron degrees of freedom), and
therefore in this paper we provide results only for 4-dim
KS vector sets of yes-no questions (KS sets for short). 
The first experiments and their designs
\cite{simon-zeil00,michler-zeil-00,cabell-01,ks-exp-03,h-rauch06}
were not literal KS sets. They made use of state-dependent
vector orientations that were additionally ``translated'' into
new measurable observables according to ingenious keys found by
their authors, because they could not be
directly implemented by reading off the orientations
of the vectors from the original set. The most recent
designs and experiments
\cite {cabello-fillip-rauch-08,b-rauch-09,k-cabello-blatt-09,amselem-cabello-09-arXiv,liu-09,moussa-09-arXiv}
dispense with state-dependent vectors.

The configurations of Hilbert space vectors and 
subspaces in KS sets have interesting symmetries 
that have intrigued many authors since the very 
discovery of the KS theorem. The set of Kochen
and Specker themselves \cite{koch-speck} is a highly
symmetrical structure that consists of three identical
substructures each of which consists of five hexagons.
This perfect symmetry enabled Kochen and Specker to ``see''
and prove their theorem. Other highly symmetrical
3-dimensional constructions were given by  Peres
\cite{peres} and  Penrose \cite{penrose-02}.
Both constructions can be given a straightforward
physical interpretation (by means of either rays in a 3-dim
spin-1 Hilbert space---equivalent to those in Euclidean
space---or by means of Majorana spin representation) and an
appealing geometrical visualisation on a cube \cite{peres-book}
or on Escher's Waterfall ornament \cite{penrose-02}.

Similar symmetries were found for higher spins, i.e., higher
dimensions. In four dimensions, even more symmetries have
been found.  Peres has found a highly symmetrical
$24\time24$ system of 24 vectors grouped in 24 tetrads
each consisting of 4 mutually orthogonal vectors.\
\cite{peres} This system can be seen as a geometrical
representation of Mermin's elegant set for a
pair of two spin-$\frac{1}{2}$ subsystems
($\frac{1}{2}\otimes\frac{1}{2}$) \cite{mermin90}
which has recently been experimentally realised
\cite{k-cabello-blatt-09}. Another representation,
based on a dodecahedron (consisting of 12 pentagons and
containing 40 rays), has been given by  Penrose.\
\cite{penrose-02,zimba-penrose,massad-arravind99}
Its physical representation is based on a Majorana
representation of a pair of entangled
spin-$\frac{3}{2}$ systems. In these examples, a
geometric visualisation is not as direct as in the
aforementioned 3-dim cases, where we can make use of a
Euclidean space instead of a Hilbert space. Nevertheless it
can help us find appropriate experimental sets even
in cases with a much higher number of rays, corresponding
to the number of measurements and preparations of a
system or the number of gates we pass the system
through---depending on the kind of experiment.

For instance, KS sets recently considered by Aravind and 
Lee-Elkin are based on the geometry of two 4-dimensional 
polytopes, the 600-cell (each cell congruent
to a regular tetrahedron) and the 120-cell,
which provide us with highly symmetrical configurations of
60 and 300 KS
rays, respectively.\ \cite{aravind-600} These highly
symmetrical structures (usually a particular regular group)
can be extended to higher dimensions \cite{ruuge05}, but
they also contain many KS substructures.
For example, the smallest 4-dim 18-9 system found by
Cabello, Estebaranz, and
Garc{\'\i}a-Alcaine \cite{cabell-est-96a}, a 20-11
found by \cite{kern}, and a number of systems with 19 through
24 rays and 10 through 24 tetrads found by Pavi\v ci\'c, 
Merlet, McKay, and Megill \cite{pmmm04a-arXiv} were all found 
to be contained in Peres' 24-24 system \cite{mpglasgow04-arXiv-0}.

These findings motivated us to find out how many
possible KS sets there are, how many of them are
sub-sets of larger ones, and whether we
can generate them from each other. This is, however,
a rather complex task which cannot be carried out as a
straightforward counting, simply because all today's
clusters and grids together would take many ages of
the Universe to carry it out using a brute force approach.
The complexity of the direct approach is illustrated,
e.g., by the fact that it took seven years before Gould
and Aravind \cite{peres-penrose-gould-aravind09}
succeeded in comparing just two of such
systems---the aforementioned Peres' and
Penrose's 3-dim KS systems---and proving them
isomorphic to each other.
Instead, we developed a new way of
describing and visualising KS systems, wrote many
new algorithms and programs, and discovered many new
symmetries and other features of the systems.

In Ref.\ \cite{pmmm04a-arXiv} we gave algorithms
for exhaustive generation of KS system containing arbitrary
number of vectors with all possible numbers of their blocks
in any number of dimensions with vector component values
from any set. Then we scanned all the systems with up to
(but not included) 23 vectors in a 4-dim space. In the
meantime, we also scanned all the systems with 23 vectors,
and the results revealed many new features.
In particular, it turned out that all possible KS vector
sets with up to and including 23 vectors and with components
from the set \{-1,0,1\} are contained in Peres'
24-24 set. It was obvious that by simply pealing off
blocks of vectors from the 24-24 set we get many more
subsets. But then we discovered that by using a lattice
representation of vector rays and filtering them
through dispersive states that we can define on these lattices,
we get exactly the KS subsets contained in the 24-24 set.
Via this method, we get additional 844 24-vector systems contained
in the 24-24 set. We conjecture that these are all
possible KS sets with 24 vectors with components
from the set \{-1,0,1\}.

We also found 37 new KS sets with 22 through 24 vectors
with component values from other sets (not from \{-1,0,1\}).

\section{Algorithms}
\label{sec:algo}

To obtain our results we used the algorithms that are
described in detail in \cite{pmmm04a-arXiv} and some
others that we describe in the Appendix A.

We start by describing vectors as vertices (points)
and orthogonalities between them as edges (lines connecting
vertices), thus obtaining MMP diagrams
\cite{mporl02-arXiv,mpglasgow04-arXiv-0,bdm-ndm-mp-1} 
which are defined as follows:
\begin{itemize}
\item[1.]Every vertex belongs to at least one edge;
\item[2.]Every edge contains at least 3 vertices;
\item[3.]Edges that intersect each other in $n-2$
         vertices contain at least $n$ vertices;
\end{itemize}
We denote vertices of MMP diagrams by {\tt 1,2,..,A,B,..a,b,..}.
There is no upper limit for the number of vertices and/or
edges in our algorithms and/or programs.

Isomorphism-free generation of MMP diagrams follows the general
principles established by~\cite{mckay98}, which we now recount briefly.
Deleting an edge from an MMP diagram, together with any vertices that
lie only on that edge, yields another MMP diagram (perhaps the vacuous
one with no vertices).  Consequently, every MMP diagram can
be constructed by starting with the vacuous diagram and adding one edge
at a time, at each stage obtaining a new MMP diagram.
We can represent this process as a rooted tree whose vertices correspond
to MMP diagrams, in which the vertices and edges have unique labels.
The vacuous
diagram is at the root of the tree, and for any other diagram its parent
node is the diagram formed by deleting the edge with the highest label.
The isomorph rejection problem is to prune this tree until it contains
just one representative of each isomorphism class of diagram.

To find diagrams that cannot be ascribed {\tt 0-1} values,
we apply an algorithm which we call {\tt states01} and which is
based on the lattice theory of Hilbert space states.
The algorithm is an exhaustive search of MMP diagrams with backtracking.
The criterion for assigning {\tt 0-1} (dispersion-free) states
is that each edge must contain exactly one vertex assigned to 1, with
the others assigned to 0. As soon as a vertex on an edge is assigned a
1, all other vertices on that edge become constrained to 0, and so on.

\section{Results}
\label{sec:res}

To find KS vectors, we follow the idea put forward in
\cite{mporl02-arXiv,mpglasgow04-arXiv-0} and proceed so as to require that
their number, i.e.~the number of vertices within edges,
corresponds to the dimension of the experimental space
${\mathbb R}^n$ and that edges
correspond to $n(n-1)/2$ equations resulting from inner products
of vectors being equal to zero (meaning
orthogonality). So, e.g., an edge of length 4, {\tt BCDE},
represents the following 6 equations:
\begin{eqnarray}
&&{\mathbf a}_B\cdot{\mathbf a}_C=
a_{B1}a_{C1}+a_{B2}a_{C2}+a_{B3}a_{C3}+a_{B4}a_{C4}=0,\nonumber\\
&&{\mathbf a}_B\cdot{\mathbf a}_D=
a_{B1}a_{D1}+a_{B2}a_{D2}+a_{B3}a_{D3}+a_{B4}a_{D4}=0,\nonumber\\
&&{\mathbf a}_B\cdot{\mathbf a}_E=
a_{B1}a_{E1}+a_{B2}a_{E2}+a_{B3}a_{E3}+a_{B4}a_{E4}=0,\nonumber\\
&&{\mathbf a}_C\cdot{\mathbf a}_D=
a_{C1}a_{D1}+a_{C2}a_{D2}+a_{C3}a_{D3}+a_{C4}a_{D4}=0,\nonumber\\
&&{\mathbf a}_C\cdot{\mathbf a}_E=
a_{C1}a_{E1}+a_{C2}a_{E2}+a_{C3}a_{E3}+a_{C4}a_{E4}=0,\nonumber\\
&&{\mathbf a}_D\cdot{\mathbf a}_E=
a_{D1}a_{E1}+a_{D2}a_{E2}+a_{D3}a_{E3}+a_{D4}a_{E4}=0.
\label{eq:mmp-eq}
\end{eqnarray}

Each possible combination of edges for a chosen number of vertices
corresponds to a system of such nonlinear equations. A solution
to systems which correspond to MMP diagrams without {\tt 0-1}
states is a set of components of KS vectors we want to
find. Thus the main method for finding {\em all} KS vectors is to
exhaustively generate all MMP diagrams, then pick out all
those diagrams that cannot have {\tt 0-1} states, then establish
the correspondence between the latter diagrams and the equations
for the vectors as shown  in Eq.~(\ref{eq:mmp-eq}), and finally
solve the systems of the so obtained equations.

To find solutions in the set \{-1,0,1\} we use the program
{\tt vectorfind}, and to find solutions in the set of real
numbers we use the interval analysis as described in detail
in \cite{pmmm04a-arXiv,pmmm04b-web}. There is no other upper
limit for the number of vertices and edges of the generated
MMP diagrams and solved equations apart from the computational
power of today's supercomputers.

In Table \ref{T:tableA} we give the numbers of all
18-~through 24-vector sets with component values from
\{-1,0,1\} that we generated and solved with the help of the
aforementioned algorithms and programs. Vector sets with
vector component values from other sets then  \{-1,0,1\}
are given at the and of the paper.

\font\1=cmss8
\begin{table}[h]
\setlength{\tabcolsep}{4.1pt}
{\small\begin{center}
\begin{tabular}{|r|rrrrrrrrrrrrrrrr|r|}
\hline
$\setminus$&9&10&11&12&13&14&15&16&17&18&19&20&21&22&23&24&\it total\\
\hline
18&1&&&&&&&&&&&&&&&&1\\
19&&1&&&&&&&&&&&&&&&1\\
20&&1&5&\ 1&&&&&&&&&&&&&7\\
21&&&2&11&4&1&&&&&&&&&&&18\\
22&&&1&9&36&23&12&3&1&&&&&&&&85\\
23&&&&2&19&76&79&58&27&11&3&1&&&&&276\\
24&&&&1&6&39&137&187&188&136&83&41&18&6&2&1&845\\
\hline
\it total&1&2&8&24&65&139&228&248&216&147&86&42&18&6&2&1&1233\\
\hline
\end{tabular}
\end{center}}
\caption{KS sets for systems with 4 degrees of freedom
with up to 24 vectors with component values from \{-1,0,1\}.}
\label{T:tableA}
\end{table}

We reported on the properties of the KS sets with 18 through (including) 22
vectors in \cite{pmmm04a-arXiv}.\footnote{Notice that here (as opposed to
\cite{pmmm04a-arXiv}) the sets with loops of size 2 and 3 are put together.}
It took two weeks on our cluster with 500 3.4 GHz processors
(recalculated) in 2004. For the present results, we ran a parallel
computation for 23-vector sets and obtained the 275 sets given
in the 6th row of Table \ref{T:tableA}. This took about two months
on our cluster. We then analysed the data and conjectured that
all sets with solutions from \{-1,0,1\} might be
subsets of the aforementioned 24-24 set. Our program
{\tt subgraph} confirmed the conjecture.

That meant that we can actually get all 18- through 24-vector sets by
stripping vectors and tetrads---vertices and edges in MMP notation---of
the 24-24 set and filtering it with our {\tt state01} program described in
\cite{pmmm04a-arXiv}. We wrote the program {\tt subset} to generate all subsets
(i.e. MMP diagrams with edges removed) of the 24-24 set.  From these, we
determined the ones with 18 through 23 vectors that are isomorphic with
the ones we previously obtained on our cluster. It is interesting that
all such stripped sets filtered by {\tt state01} have solutions.
In addition,  we determined (again filtering the output of {\tt subset})
844 24-vector sets with 12 through 23 tetrads (MMP diagrams with 24
vertices with 12 through 23 edges). They are given in the seventh row
of Table \ref{T:tableA}. All that, i.e., obtaining all 1232 sets shown
in Table \ref{T:tableA} with their vector component values from the
24-24-set, took a few minutes on a single PC.

In that way we can even get new sets with up to 41 vectors (upper limit
for the solutions from \{-1,0,1\}  \cite{pmmm04a-arXiv}) simply by adding
new vectors and tetrads to the sets from the 7th row of  Table \ref{T:tableA}

For a higher number of
vertices we might find KS sets that do not contain any of the sets from
Table \ref{T:tableA} as their subsets. If their vectors had their
component values from the set \{-1,0,1\}, they should have loops of
order higher than six because they should not have any of the above
1,231 sets as their subsets. With today's computer power, such a
search is not feasible, though.

\begin{figure}[hbt]
\begin{center}
\includegraphics[width=\textwidth]{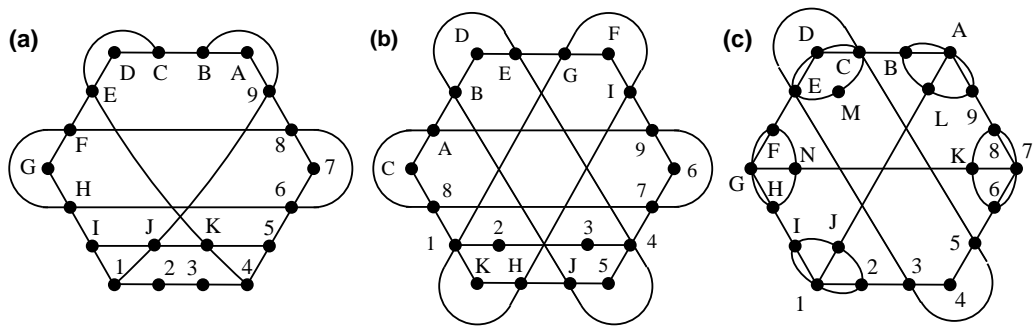}
\end{center}
\caption{KS sets: (a) 20-10; (b) the same 20-10 but redrawn so as to
match the visual appearance of 18-9 in Fig.\ 1 from \cite{cabello-08}
and Fig.\ 3 of  \cite{pmmm04a-arXiv}; (c) 23-14 which contains
neither 18-19 nor (a), (b), (c) from  Fig.\ 3 of  \cite{pmmm04a-arXiv}.}
\label{fig:solutions1}
\end{figure}

We analysed the obtained vector sets and obtained the properties we present
below. All the vector sets contain a hexagon MMP loop
{\tt 1234,4567, 789A,ABCD,DEFG,GHI1} which is always given in our figures
[except in Fig.\ \ref{fig:solutions1} (b)] and
for which we assume it is present
whenever we give a new KS set. For instance, for 20-10
from Fig.\ \ref{fig:solutions1} (a)
we just write: {\tt H68F,IJK5,1J9B,4KEC}.

The set 20-10 contains the smallest system 18-9. To determine the
orientations of its vectors, we use the program {\tt vectorfind}. 
It gives the component values given in Table \ref{T:tableB} of 
\ref{sec:add}.

Previously, we found  two smallest (20-11) KS sets
that do not contain the smallest
18-9 set (Fig.\ 4 (a) and (b) of \cite{pmmm04a-arXiv}) and two smallest
(22-13) sets that contain neither of the previous sets
(Fig.\ 4 (c) and (d) of
\cite{pmmm04a-arXiv}).

Our new results show that there are two 23-14s that contain neither the
above 18-9, nor the two 20-11s, nor the first of the above 22-13s. One of
them, {\tt 12JI,1JLA,35CE,678K,9ABL,CDEM,FGHN,GNK7}, is given in Fig.\
\ref{fig:solutions1} (c). It contains  (d) from  Fig.\ 3 of  \cite{pmmm04a-arXiv}.

\begin{figure}[hbt]
\begin{center}
\includegraphics[width=\textwidth]{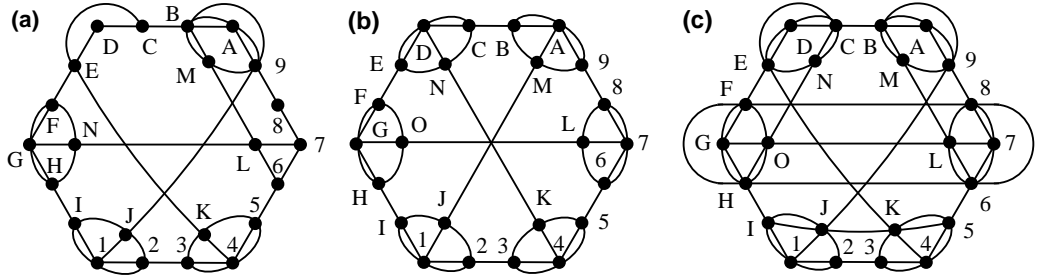}
\end{center}
\caption{(a) 23-14 which contains neither 18-19 nor (a), (b), (d) from
Fig.\ 3 of  \cite{pmmm04a-arXiv};  (b) 24-15 set (the only one that exists) that
does not contain any of the previous sets; (c) 24-20 that contains all previous sets; }
\label{fig:solutions2}
\end{figure}

There are also two 23-14s that contain neither the above 18-9, nor the two 20-11s,
nor the second of the above 22-13s. One of them,
{\tt 12JI,1J9B,345K, 4KEC,6LMB,9ABM,FGHN,GNL7}, is given in
Fig.\ \ref{fig:solutions2} (a).
It contains  (c) from  Fig.\ 3 of  \cite{pmmm04a-arXiv}.

The vectors component values for the two KS sets are given in
Table \ref{T:tableB} of \ref{sec:add}.

In Fig.\ \ref{fig:solutions2} (b) we give the only set (24-15)
that does not contain any of the previous sets. The set (c) is the one which
contains all the previous sets. Their MMP notations can easily be read off
their figures.

The vector component values for the two 23-14 KS sets are
given in Table \ref{T:tableB} of \ref{sec:add}.

Additional KS sets are given in Appendix B.

KS sets with vectors having component values from sets other than
\{-1,0,1\} are less numerous then the ones with values from
\{-1,0,1\}. They are not our primary target in this paper and
we shall present only several examples below while the
exhaustive generation of these sets is under way.\ \cite{pmm09}

All 37 KS sets with 22 through 24 vectors with component values from
sets other than \{-1,0,1\} would have component values from \{-1,0,1\}
if we discarded vectors that share only one tetrad. But we clearly
cannot do so because we have to have all vectors in every tetrad
to be able to assign 1 to one and 0 to three of them. This confirms
the results obtained in \cite{pmmm04a-arXiv,larsson-arXiv}.

\begin{figure}[hbt]
\begin{center}
\includegraphics[width=\textwidth]{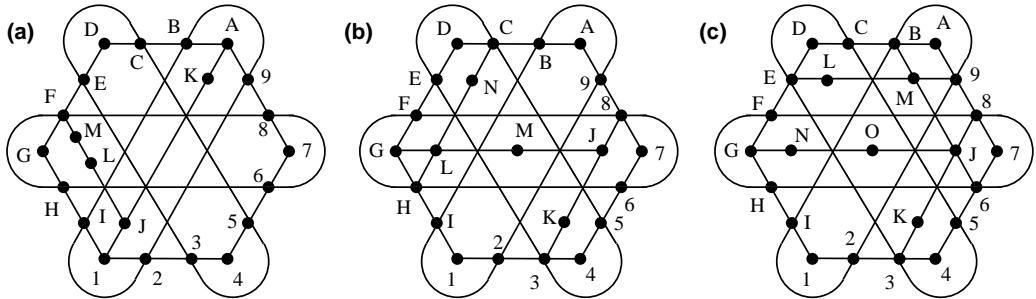}
\end{center}
\caption{KS sets with vectors whose components are not from  \{-1,0,1\}:
(a) 22-11; (b) 23-12; (c) 24-14.}
\label{fig:solutions3}
\end{figure}

All of these sets contain the 18-9 set. The smallest one is 22-11:
{\tt 25BE,1AJK, JFLM,68FH,39IC.} shown in Fig.\ \ref{fig:solutions3} (a)
It contains 20-10 from Fig.\ \ref{fig:solutions1} (a).

The vector component values for this KS set are
given in Table \ref{T:tableB} of \ref{sec:add}.

23-12 KS set shown in Fig.\ \ref{fig:solutions3} (b) contains
the 18-9, the 20-10, and the 22-11. And 24-14 set shown in  Fig.\
\ref{fig:solutions3} (c) contains the 18-9, the 20-10, a 21-11,
the 22-11, a 22-12, a 22-13, and a 23-12. Additional such KS sets
and vector components the reader can find in \cite{pmm09}.)

\section{Conclusions}
\label{sec:con}

We sum up our results as follows. All possible 388 KS sets
for systems with 4 degrees of freedom with 18 through 23 vectors
and 844 KS sets with 24 vectors with component values from
\{-1,0,1\} can be obtained by ``peeling'' vectors off a single system
provided---in effect---by  Peres 20 years ago. But we would not
know that the sets with  18 through 23 vectors obtained by
such peeling exhaust all possible KS sets up to 23 vectors
without extensive computation we carried out. And the computation
would not have been feasible without putting together the theory
of hypergraphs, lattice theory, and interval analysis, and many
algorithms and programs we devised for the purpose.

Among particular features of KS sets we presented in Section
\ref{sec:res}, we would like to single out the one about the
so-called {\em critical sets}, i.e., those KS sets that do not
properly contain any KS subset.\ \cite{zimba-penrose,aravind-600}
We found out that there are altogether six critical subsets
of Peres' 24-24 set. These are
18-9,\cite{cabell-est-96a} 20-11,\cite{kern}
another 20-11 and two 22-13s,\cite{pmmm04a-arXiv} and 24-15
[given in Fig.\ \ref{fig:solutions2}(b)].

There exist sets with 22 and more vectors with component values that are
not from \{-1,0,1\} and that are not isomorphic to any of the 1,233 sets
mentioned above.  Unlike the ``$\{$-1,0,1$\}$ sets,'' they can be obtained
only by extensive generation of MMP diagrams and computation of their
properties, which we are currently carrying out.\ \cite{pmm09}

As a final note, we mention that all 4-dim KS sets we have considered
contain a single hexagon as the biggest loop formed by their tetrads.  A
geometrical interpretation of this fact is an open question, because in
general there is no particular limit on the loop size in non-KS
orthogonal tetrads of rays contained in 4-dim Hilbert space sets.

\bigskip\bigskip
\noindent{{\bf Acknowledgements}

\medskip
Supported by the {\em Ministry of Science, Education, and
Sport of Croatia} through {\em Distributed Processing and
Scientific Data Visualization} program and {\em Quantum
Computation: Parallelism and Visualization} project
(082-0982562-3160).

\smallskip
Computational support was provided by the cluster {\em Isabella} of 
the {\em University Computing Centre} of the 
{\em University of Zagreb} and by the {\em Croatian National Grid 
Infrastructure}.}

\bigskip\bigskip

\appendix
\section{Programs}
\label{sec:app}

The primary programs used for this project (and referenced earlier 
in this paper) are {\tt subgraph}, {\tt states01}, {\tt vectorfind}, 
and {\tt subset}.\footnote{They can be downloaded from 
{\tt http://us.metamath.org/\#ql} or from 
\hbox to 50pt{\hfill} \ \hbox to 18pt{\hfill}
 {\tt http://m3k.grad.hr/ql}.} Each is a stand-alone 
{\sc ANSII} C program.  Each program typically has several input, 
output, and other options, which can be listed with the {\tt --help} 
option by typing e.g.\ {\tt subgraph --help} at the Unix or Linux 
command-line prompt.  Below we describe their main algorithms.

\subsection{{\tt subgraph}}

The program {\tt subgraph} takes as its input two hypergraphs in the
form of MMP diagrams, a test graph and a reference graph.  It will
indicate whether or not the test graph is a subgraph of the reference
graph, using the following algorithm (suggested by Brendan McKay).

Let the edges of the test graph be $w_1,w_2,\ldots,w_m$ and of the
reference graph $v_1,v_2,\ldots,v_n$, where $m\le n$.  (If $m>n$, the
subgraph relation will fail.)  The problem is to find a sequence
$v'_1,\ldots,v'_m \subseteq \{v_1,v_2,\ldots,v_n\}$ such that the
mapping $w_i \mapsto v'_i$ ($i=1.\ldots.m$) is an isomorphism.  We
construct this sequence one element at a time.  For $v'_1$, we choose an
element from $\{v_1,\ldots,v_n\}$.  For $v'_2$, we choose an element
from $\{v_1,\ldots,v_n\}$ whose relationship (described below) to $v'_1$
is the same as the relationship of $w_2$ to $w_1$.  If there is no such
$v'_2$, we backtrack and choose another $v'_1$.  Next, $v'_3$ is
anything whose relationship to $v'_1,v'_2$ is the same as the relationship
of $w_3$ to $w_1,w_2$.  And so on, backtracking recursively if necessary
until a $v'_m$ is found.  If this process is successful, the subgraph
relation holds; if, on the other hand, the backtracking is exhausted,
the subgraph relation fails.

We determine the condition ``$v'_{k+1}$ is in the same relationship to
$v'_1,\ldots,v'_k$ as $w_{k+1}$ is to $w_1,\ldots,w_k$'' as follows.
Suppose we have chosen $v'_1,\ldots,v'_k$ and wish to find the
possibilities for $v'_{k+1}$.  We look at at $w_{k+1}$:  it has
(e.g.~for a 4-dim hypergraph) 4 vertices, and each is a member of some
(possibly none) of the edges $w_1,\ldots,w[k]$.  So $w_{k+1}$ gives a
set of 4 subsets (in terms of indices) of $\{1,2,\ldots,k\}$.
Similarly, any edge $x$ in the reference graph gives a set of 4 subsets
of $\{1,2,\ldots,k\}$ that say which of the edges $v'_1,\ldots,v'_k$
contain each of the 4 vertices of $x$.  The choices for $v'_{k+1}$ are
those edges of the reference graph which give the same 4 subsets of
$\{1,2,\ldots,k\}$ as $w_{k+1}$ gives (and of course $v'_{k+1}$ is
different from $v'_1,\ldots,v'_k$).

\subsection{{\tt states01}}

The program {\tt states01} indicates whether or not an MMP diagram can
be assigned a non-dispersive state, in other words whether there exists
a 0-1 assignment to all vertices such that each edge contains exactly
one vertex assigned with 1. The program does an exhaustive search of all
possible assignments to the MMP diagram, using a fast backtracking
algorithm.  More details are described in Ref.~\cite{pmmm04a-arXiv}.

\subsection{{\tt vectorfind}}

The program {\tt vectorfind} takes as its input an MMP diagram supplied
by the user.  It attempts to assign to each vertex a 3-dim or 4-dim
vector (when the MMP diagram has 3 or 4 vertices per edge respectively),
such that the following constraints are satisfied:  (1) each vector,
chosen from a predetermined set specified by the user, must be unique
(non-parallel to all the others), and (2) the vectors assigned to the
vertices in a given edge must be mutually orthogonal (i.e. have inner
product equal to zero).  If an assignment is found, it is printed out;
otherwise the failure to find one is indicated.  The algorithm is an
exhaustive search of all possible assignments from the used-specified
vector set, using recursive backtracking.

A goal of the algorithm is to achieve extremely fast run time (compared
to the more general interval analysis method described in Section
\ref{sec:res}). While worst-case run time can grow exponentially with
the number of vertices, typically the program's speed is much faster.
An internal optimisation processes vertices with the most edges before
others to encourage early backtracking in the recursive search.
A user-settable timeout will abandon the relatively rare attempts that
take too long (and likely don't have a solution).  For the standard K-S
sets in the literature, vector assignments are found almost
instantaneously on a desktop computer.

\subsection{{\tt subset}}

The program {\tt subset} is a relatively simple utility program that
generates all subsets of the set of edges of its input MMP diagram.  By
default, subsets containing isolated edges (ones not connected to any
other edge) are suppressed.  The output will consist of $2^{n-1}$ MMP
diagrams minus the suppressed ones, where $n$ is the number of edges.
The program does not check for isomorphisms, so it is possible that some
of its output diagrams are isomorphic to each other.  (The program {\tt
subgraph} is one way to filter these if desired.)

\bigskip\bigskip
\section{Additional Results}
\label{sec:add}

\bigskip

Using our program {\tt subgraph} and several ad hoc Linux scripts
for collecting and filtering outputs we obtained the following results.
The complete encoding of the KS sets we use below include hexagons
which we only assumed in our figure representations above.

\font\stt=cmtt10

{\stt KLMN,GHIJ,DEFJ,BCFI,9ABC,78DE,56GH,1234,34AC,248E,146H,9CMN,7ELN,5HLM.}\break
and
{\stt KLMN,HIJN,DEFG,9ABC,5678,234J,178I,1BCH,4FGN,68EG,ACDG,23LM,358M,39CL.}\break
are the other two 23-14 sets that do not contain Fig.\ 3(c) and
Fig.\ 3(d) of  \cite{pmmm04a-arXiv}, respectively and that do
contain Fig.\ 3(d) and  Fig.\ 3(c), respectively [we call them (c) and (d)
below].

\begin{table}[!b]
\setlength{\tabcolsep}{2.9pt}
{\small\begin{center}
\begin{tabular}{|r|r|r|r|r|r|r|}
\hline
&{\hbox to 33pt{18-9\hfill}}&{\hbox to 33pt{23-14a\hfill}}&{\hbox to 33pt{23-14b\hfill}}&{\hbox to 33pt{24-15\hfill}}&{\hbox to 33pt{24-20\hfill}}&{\hbox to 40pt{22-11\hfill}}\\
\hline
1&\{1,0,0,1\}&\{0,0,0,1\}&\{0,0,0,1\}&\{0,0,0,1\}&\{0,0,0,1\}&\{0,1,0,0\}\\
2&\{0,1,0,0\}&\{0,0,1,0\}&\{1,0,0,0\}&\{1,0,0,0\}&\{0,1,1,0\}&\{0,0,1,0\}\\
3&\{0,0,1,0\}&\{1,-1,0,0\}&\{0,1,1,0\}&\{0,1,1,0\}&\{1,0,0,0\}&\{1,0,0,0\}\\
4&\{1,0,0,-1\}&\{1,1,0,0\}&\{0,1,-1,0\}&\{0,1,-1,0\}&\{0,1,-1,0\}&\{0,0,0,1\}\\
5&\{1,0,1,0\}&\{0,0,1,1\}&\{1,0,0,-1\}&\{1,0,0,-1\}&\{1,1,1,1\}&\{$\frac{1}{\sqrt{2}}$,$\frac{1}{\sqrt{2}}$,0,0\}\\
6&\{1,-1,-1,1\}&\{1,-1,1,-1\}&\{1,1,1,1\}&\{1,1,1,1\}&\{1,0,0,-1\}&\{$\frac{1}{2}$,-$\frac{1}{2}$,-$\frac{1}{\sqrt{2}}$,0\}\\
7&\{1,-1,-1,-1\}&\{1,-1,-1,1\}&\{1,-1,-1,1\}&\{1,-1,-1,1\}&\{1,-1,-1,1\}&\{$\frac{1}{2}$,-$\frac{1}{2}$,$\frac{1}{\sqrt{2}}$,0\}\\
8&\{1,1,0,0\}&\{1,1,1,1\}&\{1,-1,1,-1\}&\{1,1,-1,-1\}&\{1,0,1,0\}&\{$\frac{1}{2}$,$\frac{1}{2}$,0,-$\frac{1}{\sqrt{2}}$\}\\
9&\{1,-1,0,0\}&\{1,0,0,-1\}&\{1,1,0,0\}&\{0,1,0,1\}&\{1,1,-1,-1\}&\{0,$\frac{1}{\sqrt{2}}$,$\frac{1}{2}$,$\frac{1}{2}$\}\\
A&\{1,1,1,1\}&\{0,1,-1,0\}&\{0,0,1,1\}&\{1,0,1,0\}&\{0,1,0,1\}&\{$\frac{1}{\sqrt{2}}$,0,-$\frac{1}{2}$,$\frac{1}{2}$\}\\
B&\{1,1,-1,-1\}&\{1,0,0,1\}&\{1,-1,0,0\}&\{0,1,0,-1\}&\{1,0,-1,0\}&\{$\frac{1}{2}$,-$\frac{1}{2}$,0,-$\frac{1}{\sqrt{2}}$\}\\
C&\{0,0,1,-1\}&\{1,1,1,-1\}&\{1,1,1,-1\}&\{1,1,-1,1\}&\{1,1,1,-1\}&\{0,-$\frac{1}{\sqrt{2}}$,$\frac{1}{2}$,$\frac{1}{2}$\}\\
D&\{1,1,1,-1\}&\{1,-1,-1,-1\}&\{1,1,-1,1\}&\{1,-1,-1,-1\}&\{1,-1,1,1\}&\{$\frac{1}{2}$,$\frac{1}{2}$,$\frac{1}{\sqrt{2}}$,0\}\\
E&\{1,0,-1,0\}&\{1,1,-1,1\}&\{1,-1,-1,-1\}&\{1,-1,1,1\}&\{0,0,1,-1\}&\{$\frac{1}{2}$,-$\frac{1}{2}$,0,$\frac{1}{\sqrt{2}}$\}\\
F&\{0,1,0,1\}&\{0,1,0,-1\}&\{0,1,0,-1\}&\{0,0,1,-1\}&\{1,-1,-1,-1\}&\{0,$\frac{1}{\sqrt{2}}$,-$\frac{1}{2}$,$\frac{1}{2}$\}\\
G&\{1,-1,1,1\}&\{1,0,1,0\}&\{1,0,1,0\}&\{1,1,0,0\}&\{1,1,0,0\}&\{$\frac{1}{\sqrt{2}}$,0,-$\frac{1}{2}$,-$\frac{1}{2}$\}\\
H&\{0,0,0,1\}&\{1,0,-1,0\}&\{1,0,-1,0\}&\{1,-1,0,0\}&\{1,-1,0,0\}&\{$\frac{1}{\sqrt{2}}$,0,$\frac{1}{2}$,$\frac{1}{2}$\}\\
I&\{0,1,-1,0\}&\{0,1,0,0\}&\{0,1,0,0\}&\{0,0,1,0\}&\{0,0,1,0\}&\{0,0,-$\frac{1}{\sqrt{2}}$,$\frac{1}{\sqrt{2}}$\}\\
J&&\{1,0,0,0\}&\{0,0,1,0\}&\{0,0,1,1\}&\{0,1,0,0\}&\{0,0,$\frac{1}{\sqrt{2}}$,-$\frac{1}{\sqrt{2}}$\}\\
K&&\{1,1,-1,-1\}&\{1,0,0,1\}&\{1,0,0,1\}&\{1,0,0,1\}&\{-$\frac{1}{\sqrt{2}}$,0,-$\frac{1}{2}$,$\frac{1}{2}$\}\\
L&&\{0,1,1,0\}&\{1,1,-1,-1\}&\{1,-1,1,-1\}&\{1,1,-1,1\}&\{$\frac{\sqrt{3}}{2}$,-$\frac{\sqrt{2}}{4}$,-$\frac{1}{4}$,$\frac{1}{4}$\}$\!\!$\\
M&&\{1,-1,1,1\}&\{0,0,1,-1\}&\{1,0,-1,0\}&\{1,-1,1,-1\}&\{$\frac{1}{2}$,$\frac{\sqrt{3}}{2\sqrt{2}}$,$\frac{\sqrt{3}}{4}$,-$\frac{\sqrt{3}}{4}$\}$\!\!$\\
N&&\{0,1,0,1\}&\{0,1,0,1\}&\{1,1,1,-1\}&\{0,1,0,-1\}&\\
O&&&&\{0,1,0,0\}&\{0,0,1,1\}&\\
\hline
\end{tabular}
\end{center}}
\caption{Table of Vector Component Values for Some Chosen KS Sets}
\label{T:tableB}
\end{table}

Other KS sets that contain neither 18-9, nor the two 20-11s are the following
23-15 (1), 24-14 (1), 24-15 (10), 24-16 (5), and 24-17 (2).

23-15 contains both (c) and (d):

{\noindent\stt
KLMN,GHIJ,CDEF,ABEF,9BIJ,789A,DFMN,HJLN,3456,2568,1347,1278,46CF,45GJ,18KN.}

24-14 contains both (c) and (d):

{\noindent
{\stt LMNO,HIJK,DEFG,9ABC,5678,1234,78KO,BCJO,34IN,FGHN,2468,14AC,58EG,9CDG.}
}

24-15(2,4,6-9) do not contain (c) and 24-15-(1,3-5,10) do not contain
(d); thus 24-15-(4) is a critical KS set isomorphic to the one given
in Fig.\ \ref{fig:solutions2}(b):

{\noindent
{\stt
LMNO,HIJK,DEFG,9ABC,5678,1234,3478,24BC,14FG,68JK,ACIK,EGIJ,58NO,9CMO,DGMN.\break
LMNO,HIJK,DEFG,9ABC,5678,1234,34BC,78AC,24FG,68EG,9CJK,DGIK,14NO,58MO,IJMN.\break
LMNO,HIJK,DEFG,9ABC,5678,1234,34CK,248O,14FG,ABIJ,67MN,5BJO,79KN,BEGI,7DGM.\break
LMNO,HIJK,DEFG,9ABC,5678,1234,34FG,78EG,BCDG,24JK,68IK,ACHK,14NO,58MO,9CLO.\break
LMNO,HIJK,DEFG,9ABC,5678,1234,478K,3BCK,68FG,ACEG,12IJ,58NO,9CMO,2DGJ,2IMN.\break
LMNO,HIJK,DEFG,9ABC,5678,1234,478O,3BCO,68FG,ACEG,58JK,9CIK,12MN,2DGN,1HKN.\break
LMNO,HIJK,DEFG,9ABC,5678,1234,48KO,7FGK,3BCO,ACEG,56IJ,12MN,69CJ,2DGN,26IM.\break
LMNO,HIJK,DEFG,9ABC,5678,1234,4FGK,78EG,BCDG,3KNO,68MO,ACMN,12IJ,258J,29CI.\break
LMNO,HIJK,DEFG,9ABC,5678,34KO,78JN,BCIN,68FG,ACEG,12HM,1234,DGHO,2458,149C.\break
LMNO,IJKO,EFGH,ABCD,6789,345K,125N,289J,1CDJ,479M,3BDM,69GH,ADFH,5EHO,5KNO.}}

24-16(2,3) do not contain (c):

{\noindent
{\stt
LMNO$,$HIJK$,$DEFG$,$9ABC$,$5678$,$8BCO$,$7FGO$,$ACJK$,$EGIK$,$1234$,$346N$,$249C$,$14DG$,$5HKN$,$125M$,$56MN.\break
LMNO$,$HIJK$,$DEFG$,$9ABC$,$5678$,$BCJK$,$FGIK$,$ACNO$,$EGMO$,$3478$,$1256$,$48HK$,$269C$,$16DG$,$38LO$,$1234.\break
LMNO$,$HIJK$,$DEFG$,$9ABC$,$678C$,$FGJK$,$EGNO$,$3458$,$125B$,$5CDG$,$1267$,$349A$,$27IK$,$4AHK$,$17MO$,$3ALO.\break
LMNO$,$HIJK$,$DEFG$,$9ABC$,$BCJK$,$FGIK$,$5678$,$3478$,$1256$,$1234$,$ACNO$,$EGMO$,$489C$,$47DG$,$26HK$,$16MN.\break
LMNO$,$HIJK$,$DEFG$,$9ABC$,$CFGO$,$EGJK$,$5678$,$3478$,$1256$,$1234$,$24BO$,$9AMN$,$68DG$,$14IK$,$AHKN$,$67AM.}}

24-17 contain both (c) and (d):

\font\sstt=cmtt9

{\noindent
{\sstt
LMNO$,$HIJK$,$DEFG$,$FGJK$,$EGNO$,$9ABC$,$5678$,$3478$,$24BC$,$68IK$,$ACHK$,$58MO$,$9CLO$,$14DG$,$1256$,$139A$,$1234.\break
LMNO$,$HIJK$,$JKNO$,$DEFG$,$9ABC$,$78BC$,$56FG$,$349A$,$12DE$,$3478$,$1256$,$ACIK$,$EGHK$,$48MO$,$26LO$,$259C$,$47DG.}}

\end{document}